\documentclass[a4paper,11pt]{article}
\usepackage{amsmath,amsfonts}
\usepackage[dvips]{graphicx}
\newcommand{\tg}{\textrm{g}}
\newcommand{\dd}{\textrm{d}}
\newcommand{\pp}{\prime}
\begin{document}
\title{\bf Virial mass in DGP brane cosmology}
\author{H. R. Sepangi\footnote{Electronic address: hr-sepangi@sbu.ac.ir}\quad and\quad S. Shahidi\footnote{Electronic
address: sh.shahidi@mail.sbu.ac.ir} \\
\small Department of Physics, Shahid Beheshti University, Evin,
Tehran 19839 Iran}
\maketitle

\begin{abstract}
We study the virial mass discrepancy in the context of a DPG
brane-world scenario and show that such a framework can offer viable
explanations to account for the mass discrepancy problem. This is
done by defining a geometrical mass $\mathcal{N}$ that we prove to
be proportional to the virial mass. Estimating $\mathcal{N}$ using
observational data, we show that it behaves linearly with $r$ and
has a value of the order of $M_{200}$, pointing to a possible
resolution of the virial mass discrepancy. We also obtain the radial
velocity dispersion of galaxy clusters and show that it is
compatible with the radial velocity dispersion profile of such
clusters. This velocity dispersion profile can be used to
differentiate various models predicting the virial mass.
\end{abstract}

\section{Introduction}
The past decade has been marked by the advent of an avant garde
school of thought which has tried to address a number of important
problems and observations in standard General Relativity and
cosmology, namely the hierarchy problem; the great disparity between
the fundamental forces of nature, the observation that the universe
is undergoing and accelerated expansion relating to dark energy and
the problem of galaxy rotation curves relating to dark matter. This
school of thought is based on the assumption that our 4-dimensional
observable universe, the brane, is embedded in a higher dimensional
space, the bulk, which has the geometry of an AdS space and to which
gravitons can scape but ordinary matter cannot. The AdS nature of
the bulk space would cause gravity to become localized around the
brane and would modify the gravitational potential at short
distances. Such a scenario \cite{randall}, proposed by Randall and
Sundrum (RS), has been able to account for the hierarchy of the
fundamental forces with great success and has been generating a
myriad of other scenarios and variations. Shortly after, in a
seminal work \cite{sms}, it was shown how to project the Einstein
field equations, assumed to hold in the bulk, onto the brane. An
unprecedented number of works utilizing this idea have been
appearing ever since \cite{collins,gregory}.

In an effort to relax the restriction of an Ads  bulk and hence
allowing gravity to penetrate large distances, Dvali, Gabadadze and
Poratti (DGP) \cite{dgp,dgp1} proposed an alternative model in which
the influence of gravity on the brane is accounted for by including
an induced 4-dimensional Einstein-Hilbert term to the full action.
In this model, in contrast to the standard RS model, gravity is
modified at large distances. For a comprehensive review of this
model see \cite{Lue}. The cosmological implications of this model
was investigated  in \cite{deffayet} where it was shown that the
Friedmann equation on the brane has two branches, both reducing to
the usual FRW equation at the small Hubble radius limit. However, the
important discovery was that one of these branches predict a
self-accelerating universe at late times, consistent with the
observation that our universe is undergoing an accelerated expanding
phase. So much for the success of the DGP model, a word of caution
is in order; the theory predicts the existence of ghost-like
excitations. Many scenarios have been undertaken to explain away
such ghosts, but as yet no satisfactory solution exists. The
interested reader should consult \cite{derham,koyama} for further
insight. In this paper we do not discuss such excitations since our
aim lies in studying the virial mass discrepancy in  DGP models.

One of the interesting problems in cosmology is the calculation of
the mass of cluster of galaxies. In recent years, our ability in
performing precision measurements in observational cosmology is
developed to such an extent that we can obtain  accurate values for
the mass of individual galaxies and their velocities. We may
therefore find the total mass of cluster of galaxies in two ways,
adding up the masses of individual galaxies, or do it statistically
and use the virial theorem. Since these methods represent two
aspects of the same thing we must obtain the same result. However,
almost in all clusters the virial mass is 20-30 time greater than
$M$, the mass obtained by adding up the individual masses. Also by
Newton second law we know that the mass of a galaxy is proportional
to $rv_{tg}$ where $v_{tg}$ is the tangential velocity of a test
particle located at the distance $r$ from the center of the galaxy.
As observations have shown, the tangential velocity of a test
particle remains nearly constant at large distances from the center
of the galaxy. This is, of course, in contradiction to what
Newtonian gravity predicts. One way around this is to postulate the
existence of dark matter. However, there are geometrical approaches
to address this problem, namely to use modified Einstein field
equations, as is done in brane-world models
\cite{harko_virial,heydari-sepangi} or in modified gravity
\cite{beken,moffat}.

In this paper, we use the DGP model discussed above to explain the
the virial mass discrepancy in a geometric manner.  As we shall see
later, the linearly increasing behavior of the virial mass with
distance can be explained by taking into account the extra terms
that appear in the field equations which in turn originate from the
bulk geometry. In order to do this we must have a procedure to
obtain the total mass of the cluster. We use Jean's equation and the
observational data to obtain the total mass distribution of
clusters. We also obtain the radial velocity dispersion in this
model which suggests an alternative way to obtain the virial mass of
clusters and can be used to explain the observational data. The
question of the flat rotation curves of individual galaxies requires
a separate undertaking and will be dealt with in a future work.

\section{Einstein equations on the brane}
Let us start with the standard DGP action \cite{dgp1}
\begin{align}
  S=\frac{m_4^3}{2}\int_\mathcal{M} \dd^5 x \sqrt{-\tg}\mathcal{R}+\frac{m_3^2}{2}
  \int_{\partial\mathcal{M}} \dd^4 x\sqrt{-q}(R-2\Lambda)-
  m_4^3\int_{\partial\mathcal{M}} \dd^4 x\sqrt{-q}K
  +\mathcal{S}_m(q_{\mu\nu})+\mathcal{S}_B(\tg_{AB}),
  \label{dgp_action}
\end{align}
where $\tg_{AB}$, $\mathcal{R}$ and $\mathcal{S}_B$ are the metric,
Ricci scalar and matter action of the bulk and $q_{\mu\nu}$, $R$ and
$\mathcal{S}_m$ are those of the brane with $\Lambda$ being the
brane cosmological constant, and $m^3_4$ ($m^2_3$) is the bulk
(brane) Planck scale. The third term is the Hawking-Gibbons boundary
term \cite{hawking} and $K$ is the extrinsic curvature. After
varying the action and denoting the extra dimension by $y$, we
obtain the Einstein field equations in the bulk
\begin{align}
  &m_4^3\left(\mathcal{R}_{AB}-\frac 1 2 \tg_{AB}\mathcal{R}\right)+m_3^2\delta_A^\mu \delta_B^\nu
  \left(R_{\mu\nu}-\frac 1 2 q_{\mu\nu}R\right)\delta(y)\nonumber \\
  &=\delta_A^\mu \delta_B^\nu~(T_{\mu\nu}-m^2_3\Lambda q_{\mu\nu})\delta(y)
  +\hat{T}_{AB}, \label{enstein eq}
\end{align}
where $\hat{T}_{AB}$ ($T_{\mu\nu}$) is the bulk (brane) energy
momentum tensor for which a perfect fluid form is assumed
\begin{align}
  &\hat{T}^A_{~B}=\mbox{diag} (-\rho_{_B},p_{_B},p_{_B},p_{_B},p_5),\\
  \nonumber \\
  &T^\mu_{~\nu}=\mbox{diag} (-\rho_b,p_b,p_b,p_b,0),
\end{align}
and for the bulk metric we take
\begin{align}
  \dd s^2=-e^{\nu(r,y)}~\dd t^2+e^{\mu(r,y)}~\dd r^2+r^2\left(\dd\theta^2+\sin^2\theta~\dd\varphi^2\right)
  +e^{\lambda(r,y)}~\dd y^2. \label{bulk metric}
\end{align}
The induced metric on the brane is simply
\begin{align}
  \dd s^2=-e^{\nu_0(r)}~\dd t^2+e^{\mu_0(r)}~\dd r^2+r^2\left(\dd\theta^2+\sin^2\theta~\dd\varphi^2\right),
  \label{brane metric}
\end{align}
where $\nu_0(r)=\nu(r,0)$ etc. Assuming that the brane is at $y=0$
and noting that \cite{mohammedi}
\begin{align}
  &\frac{d f}{d y}=\dot{f}\frac{d |y|}{d y}=\dot{f}\left[2\theta(y)-1\right],\label{note1} \\
  &\frac{d^2 f}{d y^2}=\ddot{f}+2\dot{f}\delta(y),\label{note2}\\
  &\left(\frac{d f}{d y}\right)\left(\frac{d h}{d y}\right)=\dot{f}\dot{h},\label{note3}
\end{align}
where $\dot{f}=\frac{d f}{d |y|}$ and $f$, $h$ are arbitrary
functions of $|y|$, we obtain the following field equations in the
bulk
\begin{align}
  &\frac{m_4^3}{4r^2}\Bigg[\left(4 + 4r\lambda^\pp+2\lambda^{\pp\pp}r^2+\lambda^{\pp 2}r^2-4r\mu^\pp
  -\mu^\pp\lambda^\pp r^2\right)e^{-\mu}
  + \left(2\ddot{\mu}r^2+4\dot{\mu}r^2\delta(y)+\dot{\mu}^2 r^2
  -\dot{\mu}\dot{\lambda} r^2\right)e^{-\lambda}-4\Bigg]\nonumber \\
  &- \frac{m_3^2}{r^2}e^{-\mu_0}\bigg[r\mu_0^\pp+e^{\mu_0}-1\bigg]\delta(y)=
  -(\rho_b+m^2_3\Lambda)\delta(y)-\rho_{_B}
  \label{e_eq_Bu_1}, \\
  &\frac{m_4^3}{4r^2}\Bigg[\left(4+4r\nu^\pp+4r\lambda^\pp
  +\nu^\pp\lambda^\pp r^2\right)e^{-\mu}
  + \left(2\ddot{\nu}r^2+4\dot{\nu}r^2\delta(y)+\dot{\nu}^2 r^2
  -\dot{\nu}\dot{\lambda} r^2\right)e^{-\lambda}-4\Bigg]\nonumber \\
  &+ \frac{m_3^2}{r^2}e^{-\mu_0}\bigg[r\nu_0^\pp-e^{\mu_0}+1\bigg]\delta(y)=(
  p_b-m^2_3\Lambda)\delta(y)+p_{_B},
  \label{e_eq_Bu_2} \\
  &\frac{m_4^3}{4r}\Bigg[\left(2\nu^\pp-2\mu^\pp+2\lambda^\pp+2\nu^{\pp\pp}r+\nu^{\pp 2}r+2\lambda^{\pp\pp}r
  +\lambda^{\pp 2}r-\mu^\pp\nu^\pp r+\nu^\pp\lambda^\pp r-\mu^\pp\lambda^\pp r\right)e^{-\mu}\nonumber\\
  &+ \left(2\ddot{\nu}r+4\dot{\nu}r\delta(y)+\dot{\nu}^2 r
  +2\ddot{\mu}r+4\dot{\mu}r\delta(y)+\dot{\mu}^2 r
  +\dot{\nu}\dot{\mu} r-\dot{\nu}\dot{\lambda} r-\dot{\mu}\dot{\lambda} r\right)e^{-\lambda}\Bigg]\nonumber \\
  &+ \frac{m_3^2}{4r}e^{-\mu_0}\bigg[2\nu^\pp_0-2\mu^\pp_0-\nu^\pp_0 \mu^\pp_0 r+2\nu^{\pp\pp}_0r+\nu^{\pp 2}r
  \bigg]\delta(y)=(p_b-m^2_3\Lambda)\delta(y)+p_{_B}, \label{e_eq_Bu_3}\\
  &\frac{m_4^3}{4r^2}\Bigg[\left(4-4r\mu^\pp+4r\nu^\pp
  \mu^\pp r^2+\nu^{\pp 2}r^2+2\nu^{\pp\pp}r^2\right)e^{-\mu}
  + \dot{\nu}\dot{\mu}r^2 e^{-\lambda}-4\Bigg]=p_5, \label{e_eq_Bu_4}\\
  &\frac{m_4^3}{4r}\bigg(2\dot{\nu}^\pp r+\nu^\pp\dot{\nu} r -\dot{\mu}\nu^\pp r-\lambda^\pp\dot{\nu} r
  -4\dot{\mu}\bigg)\big(2\theta(y)-1\big)=0\label{e_eq_Bu_5}.
\end{align}
where a prime represents derivative with respect to $r$. Since only
$\delta(y)$ can contribute to the brane equations  we obtain the
Einstein field equations on the brane
\begin{align}
  &m_3^2 e^{-\mu_0}\left(\frac{\mu^\pp_0}{r}-\frac{1}{r^2}+\frac{e^{\mu_0}}{r^2}\right)=
  \rho_b(r)+\mathcal{U}(r)+m^2_3\Lambda, \label{e_eq_b_1}\\
  &m_3^2 e^{-\mu_0}\left(\frac{\nu^\pp_0}{r}+\frac{1}{r^2}-\frac{e^{\mu_0}}{r^2}\right)=
  p_b(r)+\mathcal{P}(r)-m^2_3\Lambda, \label{e_eq_b_2}\\
  &m_3^2 e^{-\mu_0}\left(\frac{\nu^\pp_0}{2r}-\frac{\mu^\pp_0}{2r}-\frac{\mu^\pp_0\nu^\pp_0}{4}
  +\frac{\nu_0^{\pp\pp}}{2}+\frac{\nu_0^{\pp 2}}{4}\right)=p_b(r)+\big[\mathcal{P}(r)-\mathcal{U}(r)
  \big]-m^2_3\Lambda, \label{e_eq_b_3}
\end{align}
where we have defined the induced energy density $\mathcal{U}(r)$
and pressure $\mathcal{P}(r)$  as
\begin{align}
  &\mathcal{U}(r)=m_4^3~\dot{\mu}\big|_{ y=0}~e^{-\lambda_0},\\
  &\mathcal{P}(r)=-m_4^3~\dot{\nu}\big|_{ y=0}~e^{-\lambda_0}.
\end{align}
\section{The Virial theorem}
To obtain the virial theorem in the context of the model discussed
above we use the tetrad formalism by defining the following frame of
orthonormal vectors \cite{jackson}
\begin{align}
  e^{(0)}_\rho=e^{\frac{\nu_0}{2}}\delta^0_\rho,\qquad
  e^{(1)}_\rho=e^{\frac{\mu_0}{2}}\delta^1_\rho,\qquad
  e^{(2)}_\rho=r\delta^2_\rho,\qquad
  e^{(3)}_\rho=r\sin\theta~\delta^3_\rho,\label{tetrad_co}
\end{align}
where $q^{\mu\nu}e^{(a)}_\mu e^{(b)}_\nu=\eta^{(a)(b)}$ and the
tetrad indices are surrounded by parenthesis. The 4-velocity $v^\mu$
of a typical galaxy with $v^\mu v_\mu=-1$ is written as
\begin{align}
  v^{(a)}=v^\mu e^{(a)}_\mu,\qquad a=0,1,2,3.\label{4-velocity}
\end{align}
Let us start with the Boltzmann equation in tetrad formalism. If
$f(x^\mu,v^{(a)})$ represents the distribution function of galaxies,
supposed to be made of identical and collisionless point particles,
we have \cite{jackson, lindquist}
\begin{align}
  v^{(a)} e^\rho_{(a)}\frac{\partial f}{\partial
  x^\rho}+\gamma^{(a)}_{(b)(c)}v^{(b)}v^{(c)}\frac{\partial
  f}{\partial v^{(a)}}=0,\label{bolt_eq_1}
\end{align}
where
$\gamma^{(a)}_{(b)(c)}=e^{(a)}_{\rho~;\sigma}e^\rho_{(b)}e^\sigma_{(c)}$
are the Ricci rotation coefficients. Using the brane metric
(\ref{brane metric}) and assuming that $f(x^\mu,v^{(a)})$ depends on
$r$ only, the Boltzmann equation becomes
\begin{align}
  v_r\frac{\partial f}{\partial
  r}&-\left(\frac{v^2_t}{2}\nu_0^\pp-\frac{v^2_\theta+v^2_\varphi}{r}\right)\frac{\partial
  f}{\partial v_r}-\frac{v_r}{r}\left(v_\theta \frac{\partial
  f}{\partial v_\theta}+v_\varphi \frac{\partial f}{\partial
  v_\varphi}\right) \nonumber \\
  &-\frac{e^{\frac{\mu_0}{2}}v_\varphi}{r}\cot\theta\left(v_\theta
  \frac{\partial f}{\partial v_\varphi}-v_\varphi \frac{\partial
  f}{\partial v_\theta}\right)=0.\label{bolt_eq_2}
\end{align}
where we have defined
\begin{align}
  v^{(0)}=v_t,\qquad v^{(1)}=v_r,\qquad v^{(2)}=v_\theta,\qquad
  v^{(3)}=v_\varphi.\label{velocity_rep}
\end{align}
Since our metric is spherically symmetric, the coefficient of
$\cot\theta$ must be zero in (\ref{bolt_eq_2}). Multiplying equation
(\ref{bolt_eq_2}) by $mv_r dv$, where $dv=\frac{1}{v_t}dv_r
dv_\theta dv_\varphi$ is the invariant volume element in the
velocity space and $m$ is the mass of the galaxy, and integrating
over the velocity space and assuming that $f$ vanishes sufficiently
rapidly as the velocities tend to $\pm\infty$, we obtain
\begin{align}
  r\frac{\partial}{\partial
  r}\left[\rho\left<v^2_r\right>\right]+\frac{1}{2}\rho\left[\left<v^2_t\right>+\left<v^2_r\right>\right]r\nu_0^\pp
  -\rho\left[\left<v^2_\theta\right>+\left<v^2_\varphi\right>-2\left<v^2_r\right>\right]=0,\label{bolt_eq_3}
\end{align}
where  $\rho$ is the mass density and $\left<\,\,\right>$ represents
the average values. Multiplication of equation (\ref{bolt_eq_3}) by
$4\pi r^2$ and integration over the cluster of galaxies yield
\begin{align}
  -\int_0^R 4\pi\rho\left[\left<v^2_r\right>+\left<v^2_\theta\right>+
  \left<v^2_\varphi\right>\right]r^2dr+\frac{1}{2}\int_0^R 4\pi
  r^3\rho\left[\left<v^2_t\right>+\left<v^2_r\right>\right]\frac{\partial\nu_0}{\partial
  r}dr=0.\label{bolt_eq_4}
\end{align}
We can also write equation (\ref{bolt_eq_4}) in the form
\begin{align}
  2K=\frac{1}{2}\int_0^R 4\pi
  r^3\rho\left[\left<v^2_t\right>+\left<v^2_r\right>\right]\frac{\partial\nu_0}{\partial
  r}d r,  \label{bolt_eq_5}
\end{align}
since the total kinetic energy of galaxies is defined as
\begin{align}
  K=\int_0^R 2\pi\rho\left[\left<v^2_r\right>+\left<v^2_\theta\right>+
  \left<v^2_\varphi\right>\right]r^2dr.\label{kinetic_energy}
\end{align}
To obtain the virial theorem in our model we must express the
energy-momentum tensor components in the terms of the distribution
function. This is done according to
\begin{align}
  T_{\mu\nu}=\int f m v_\mu v_\nu d v,\label{en_momentum_1}
\end{align}
which leads to
\begin{align}
  \rho_b=\rho\left<v^2_t\right>,\qquad p_b=\rho\left<v_r^2\right>=\rho\left<v_\theta^2\right>=
  \rho\left<v^2_\varphi\right>.
  \label{en_momentum_2}
\end{align}
Adding equations (\ref{e_eq_b_1}), (\ref{e_eq_b_2}) and twice of
(\ref{e_eq_b_3}) yields
\begin{align}
 m_3^2 e^{-\mu_0}\left(\frac{\nu_0^\prime}{r}-\frac{\nu_0^\prime \mu_0^\prime}{4}+
 \frac{\nu_0^{\prime\prime}}{2}+\frac{\nu_0^{\pp 2}}
 {4}\right)=\frac 1 2 \rho\left<v^2\right>+\frac 1 2 \left[3\mathcal{P}(r)-
 \mathcal{U}(r)\right]-m^2_3\Lambda, \label{bolt_eq_6}
\end{align}
where we have defined
$\left<v^2\right>=\left<v_t^2\right>+\left<v_r^2\right>+\left<v_\theta^2\right>+
\left<v_\varphi^2\right>$. For the cluster of galaxies we may assume
that $\mu(r)$ and $\nu(r)$ are small so that the quadratic terms in
equation (\ref{bolt_eq_6}) do not contribute. Also, the velocity of
galaxies are much smaller than the speed of light, so we can set
$\left<v_r^2\right>,\left<v_\theta^2\right>,
\left<v_\varphi^2\right> \ll \left<v_t^2\right> \approx 1$
\cite{harko_virial}. Taking these approximations into consideration,
equation (\ref{bolt_eq_6}) is reduced to
\begin{align}
  \rho=m_3^2\frac{1}{r^2}\frac{\dd}{\dd r}\left(r^2\nu_0^\prime\right)+2m^2_3\Lambda-
  \left[3\mathcal{P}(r)-\mathcal{U}(r)\right].  \label{bolt_eq_7}
\end{align}
Multiplying equation (\ref{bolt_eq_7}) by $r^2$ and integrating from $0$ to $r$ yields
\begin{align}
  m_3^2 r^2 \nu_0^\pp- \frac{1}{4\pi} M(r)+\frac 2 3 m^2_3\Lambda
  r^3-\frac{1}{4\pi} \mathcal{N}(r)=0,\label{bolt_eq_8}
\end{align}
where
\begin{align}
  M(r)=4\pi\int^r_0 \rho r^{\prime ^2} d r^\prime,\label{mass}
\end{align}
and
\begin{align}
  \mathcal{N}(r)=4\pi\int^r_0 \left[3\mathcal{P}(r^\pp)-\mathcal{U}(r^\pp)\right] r^{\pp 2} dr^\pp.
  \label{N_mass}
\end{align}
Again, multiplying equation (\ref{bolt_eq_8}) by $\frac{dM(r)}{r}$
and integrating from $0$ to $R$, we finally obtain the generalized
virial theorem in a DGP senario
\begin{align}
  W+2K+\frac 1 3 \Lambda I+\mathcal{W}_B=0,\label{virial_th}
\end{align}
where
\begin{align}
  &W=-\frac{1}{8\pi m^2_3} \int_0^R \frac{M(r)}{r}dM(r),\label{def_work_1}\\
  &\mathcal{W}_B=-\frac{1}{2m^2_3}\int_0^R \rho r\mathcal{N}(r) dr,\label{def_work_2}
\end{align}
and
\begin{align}
  I=\int^R_0 r^2 dM(r),\label{inertial}
\end{align}
is the moment of inertia of the system. Without the last term, this
would constitute the usual virial theorem with a cosmological
constant, first derived by Jackson \cite{jackson} using the
Boltzmann equation (\ref{bolt_eq_1}) into which the metric of the
space-time is substituted. In order to obtain a relation between the
virial mass and the extra term $\mathcal{N}(r)$ which has its
origins in the bulk, we define the following radii \cite{jackson}
\begin{align}
  &R_{_V}=\frac{M^2}{\int^{_R}_{_0} \frac{M(r)}{r}dM(r)},\label{radii_1}\\
  &R^2_{_I}=\frac{\int^{_R}_{_0} r^2 dM(r)}{M(r)},\label{radii_2}\\
  &\mathcal{R}=-\frac{1}{8\pi m^2_3} \frac{\mathcal{N}^{2}}{\mathcal{W}_B},\label{radii_3}
\end{align}
where $R_{_V}$ is the virial radius and $\mathcal{R}$ is the radius defined by the extra term $\mathcal{N}$. By
defining the virial mass as
\begin{align}
  2K=\frac{1}{8\pi m^2_3}\frac{M^2_{_V}}{R_{_V}},\label{v_mass}
\end{align}
and using the relations
\begin{align}
  W=-\frac{1}{8\pi m^2_3}\frac{M^2}{R_{_V}},\qquad I=MR^2_{_I},\label{relation}
\end{align}
the generalized virial theorem (\ref{virial_th}) can be written as
\begin{align}
  \left(\frac{M_{_V}}{M}\right)^2=1-\frac{8\pi m_3^2 \Lambda}{3}\frac{R_{_I}^2 R_{_V}}{M}+
  \left(\frac{\mathcal{N}}{M}\right)^2 \left(\frac{R_{_V}}{\mathcal{R}}\right).\label{virial_eq_1}
\end{align}
The contribution of $\Lambda$ to the mass of the galaxy is several
order of magnitude smaller than the observed mass. Also $M_{_V}$ is
much larger than $M$ for most galaxies. Therefore, we can neglect
the unity and the term involving the cosmological constant in
equation (\ref{virial_eq_1}). The virial mass in our model is then
given by
\begin{align}
M_{_V}(r) \simeq \mathcal{N}(r)
\sqrt{\frac{R_{_V}}{\mathcal{R}}}.\label{viral_eq_2}
\end{align}
As can be seen, the virial mass is proportional to an extra term
stemming from the global bulk effects.

\section{Estimating $\mathcal{N}(r)$}
In order to estimate $\mathcal{N}(r)$ we must solve the Einstein
equations for $\mathcal{P}(r)$ and $\mathcal{U}(r)$ and  obtain
$\mathcal{N}(r)$ from (\ref{N_mass}). However there is a simpler way
of doing this which we will follow. First, consider the conservation
of the right-hand side of Einstein equations
(\ref{e_eq_b_1})-(\ref{e_eq_b_3})
\begin{align}
  &\nu_0^\pp=-2~\frac{p^\pp_b+\left(\mathcal{P}^\pp+\frac{\mathcal{U}^\pp}{r}\right)}
  {(\rho_b+p_b)+(\mathcal{P}+\mathcal{U})}.\label{conserve_1}\\
  &\mathcal{U}=0.\label{conserve_2}
\end{align}
This means that we only need to calculate $\mathcal{P}(r)$. In most
clusters the majority of the baryonic mass is in the form of
intra-cluster gas. Taking this assumption into consideration and
using equations (\ref{mass}) and (\ref{N_mass}), we obtain an
expression for the total mass of the cluster
\begin{align}
  \frac{\dd M_{tot}}{\dd r}=4\pi\rho_g r^2+12\pi\mathcal{P} r^2. \label{total_mass_1}
\end{align}
Another expression can be obtained from  Jean's equation
\begin{align}
  \frac{d}{dr}\left[\rho_g\sigma^2_r\right]+\rho_g(r)\frac{d\Phi}{dr}=0,\label{jean_1}
\end{align}
where $\Phi(r)$ is the gravitational potential. We also assume that
the gas is isotropically distributed inside the cluster so that the
mass-weighted velocity dispersion in the radial and tangential
directions are equal; $\sigma_r=\sigma_{\theta,\phi}$. Assuming that
the gravitational field is weak so that $\Phi(r)$ satisfies the
poisson equation  $2m_3^2\nabla^2 \Phi\approx \rho_{tot}$, the
Jean's equation is reduced to
\begin{align}
  \frac{dp_g(r)}{dr}=-\frac{1}{8\pi m_3^2}\frac{M_{tot}}{r^2}\rho_g(r),\label{jeans_2}
\end{align}
where we have used the relation $p_g=\rho_g\sigma^2_r$. As has been
shown in \cite{reiprich}, the gas density $\rho_g$ can be fitted to
the observational data by the following radial distribution
\begin{align}
\rho_g(r)=\rho_0\left(1+\frac{r^2}{r_c^2}\right)^{-\frac{3\beta}{2}},\label{density_dist_1}
\end{align}
where $r_c$ is the core radius and $\rho_0$ and $\beta$ are cluster
independent constants. For most clusters $\beta \geq \frac 2 3$
\cite{reiprich} and therefore, in the limit $r \gg r_c$ considered
here, the gas density distribution can be written as
\begin{align}
\rho_g(r)=\rho_0 \left(\frac{r}{r_c}\right)^{-3\beta},\qquad \beta\geq \frac{2}{3}. \label{density_dist_2}
\end{align}
Moreover, we choose the following equation of state for the
intra-cluster gas \cite{reiprich}
\begin{align}
p_g(r)=\frac{k_{_B}T_g}{\mu m_p}\rho_g(r),\label{equation_state}
\end{align}
where $\mu=0.61$ is the mean atomic weight of the particles in the
cluster gas and $m_p$ is the mass of proton. With these assumptions,
equation (\ref{jeans_2}) reduces to
\begin{align}
M_{tot}(r)=8\pi m_3^2\frac{3k_{_B}T_g}{\mu m_p}\beta~r.
\label{jeans_3}
\end{align}
The contribution of the gas density to the total mass of the cluster
is very small at the boundary of the cluster, i.e. where $r \gg
r_c$, as can be seen from equations (\ref{density_dist_2}) and
(\ref{total_mass_1}). We may now calculate $\mathcal{P}(r)$ from
equations (\ref{total_mass_1}) and (\ref{jeans_3}), taking the
approximation above
\begin{align}
  \mathcal{P}(r)=2 m_3^2\frac{k_{_B}T_g}{\mu m_p}\beta~\frac{1}{r^2}, \label{P_equation}
\end{align}
and using equation (\ref{N_mass}) to obtain
\begin{align}
  \mathcal{N}(r)=M_{tot}(r)=8\pi m_3^2\frac{3k_{_B}T_g}{\mu m_p}\beta~r. \label{N_equation}
\end{align}
This equation is obviously an approximation since the contribution
of the baryonic mass is neglected. However this is no cause for
concern since the value of the baryonic mass of the clusters is
about 3 orders of magnitude smaller than its total mass. As can be
seen, $\mathcal{N}(r)\propto ~r$ and since the virial mass is
proportional to $\mathcal{N}$ and the latter is proportional to $r$,
this could offer a possible resolution to the virial mass
discrepancy in the context of DGP brane worlds.

To estimate the value of $\mathcal{N}$, we first note that $8\pi
m_3^2=G_{(4)}^{-1}=\frac{4}{3}G_{_N}^{-1}$ where $G_{_N}$ is the
gravitational constant  \cite{deffayet}. A typical value of the
temperature of a cluster gas is $k_{_B}T_g\approx 5~keV$
\cite{reiprich}. The virial radius of the cluster of galaxies is
usually assumed to be $r_{200}$, indicating the radius for which the
mass density of the cluster is about $\rho_{200}=200\rho_c$, where
$\rho_c=4.6975\times10^{-27}h_{50}^2~kg/m^3$. The virial mass of the
cluster is then estimated as $M_V=M_{200}=M(r<r_{200})$. We can
therefore define the maximum extension of the $\mathcal{N}(r)$ mass
to be the radius at which $\mathcal{P}=\rho_{200}$. From equation
(\ref{P_equation}) we have
\begin{align}
  r_{max}=4.28\beta^{\frac{1}{2}}h_{50}^{-1}\left(\frac{k_{_B}T_g}{5~KeV}\right)^\frac{1}{2}~Mpc.\label{P_final}
\end{align}
Finally, we can estimate $\mathcal{N}(r)$ from (\ref{N_equation})
\begin{align}
  \mathcal{N}(r)=32.72\times 10^{14}~\beta^{\frac{3}{2}}
  \left(\frac{k_{_B}T_g}{5~KeV}\right)^{\frac{3}{2}}~M_\odot.\label{N_final}
\end{align}
This is in agreement with observational values for the virial mass of
clusters \cite{reiprich}.
\section{Radial velocity dispersion}
\begin{figure}
\centering
  \includegraphics[scale=0.5]{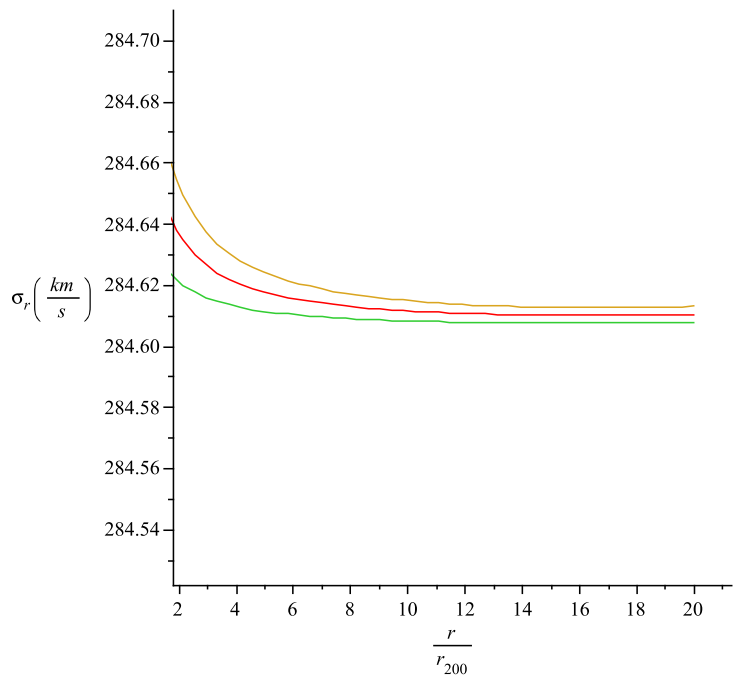}\\
  \caption{\footnotesize Radial velocity dispersion for the NGC5813 cluster with $\beta=0.766$.
  For this cluster $r_{200}=0.87Mpc$, $r_c=25Kpc$, $k_BT_g=0.52KeV$ and we have set
  $C_1=4.2,2.9,1.4~\times 10^{-8} M_\odot$ and $C_2=0.0390,0.0260,0.0130~M_\odot^2/Kpc^4$ for the top,
  middle and bottom curves respectively.}

  \label{rvd}
\end{figure}
Another important observational quantity is the radial velocity
dispersion which plays an important role in estimating the virial
mass of the clusters. As is well known, the simple form
$\sigma_r^2=B/(r+b)$ can be used to fit the observational data
\cite{carl}. The DGP model can provide an expression for the radial
velocity dispersion which we will derive in this section.

Adopting the approximations used after equation (\ref{bolt_eq_6}),
we may write equation (\ref{bolt_eq_3}) as
\begin{align}
  \frac{\dd}{\dd r}(\rho\sigma_r^2)+\frac 1 2 \rho\nu_0^\pp=0, \label{sigma_1}
\end{align}
where we have assumed that the velocity distribution in the cluster
is isotropic, so that $\left<v^2\right>=\left<v_r^2\right>
+\left<v_\theta^2\right>+\left<v_\varphi^2\right>=3\left<v_r^2\right>=3\sigma_r^2$.
Moreover, from the Einstein field equations we have
\begin{align}
  m_3^2\left(\frac{2\nu_0^\pp}{r}+\nu_0^{\pp \pp}\right)=3\mathcal{P}(r)+\rho(r). \label{sigma_2}
\end{align}
Integrating, we obtain
\begin{align}
  m_3^2 r^2 \nu_0^\pp=\frac{1}{4\pi}\mathcal{N}(r)+\frac{1}{4\pi}M(r)+C_1 , \label{sigma_3}
\end{align}
where $C_1$ is some constant. The differential form of the radial
velocity dispersion can be obtained from equations (\ref{sigma_1})
and (\ref{sigma_3})
\begin{align}
  2m_3^2 \frac{\dd}{\dd r}(\rho\sigma_r^2)=-\frac{\mathcal{N}(r)}{4\pi r^2}\rho(r)-
  \frac{M(r)}{4\pi r^2}\rho(r)-\frac{C_1}{r^2}\rho(r).
  \label{sigma_4}
\end{align}
Now, using expressions (\ref{density_dist_2}) and (\ref{N_equation})
for $\rho(r)$ and $\mathcal{N}(r)$ and by virtue of equation
(\ref{mass}) we obtain
\begin{align}
  \sigma_r^2=\frac{k_{_B}T_g}{\mu m_p}-\frac{\rho_0 r^2}{12(\beta-1)(3\beta-1)m_3^2}
  \left(\frac{r}{r_c}\right)^{-3\beta}+
  \frac{C_1}{2(3\beta+1)m_3^2}\frac{1}{r}+ \frac{C_2}{2m_3^2\rho_0}
  \left(\frac{r}{r_c}\right)^{3\beta} \qquad \beta\neq 1. \label{sigma_5}
\end{align}
For $\beta=1$ we have
\begin{align}
  \sigma_r^2=\frac{k_{_B}T_g}{\mu m_p}+\frac{\rho_0 r_c^3}{8m_3^2}\frac{\ln r}{r}+
  \frac{C_3}{r}+ \frac{C_2}{2m_3^2\rho_0}
  \left(\frac{r}{r_c}\right)^3. \qquad \label{sigma_6}
\end{align}
Our expression for $\sigma_r^2$ can therefore be used to fit the
observational data. In figure \ref{rvd} we have plotted the radial
velocity dispersion for the cluster NGC5813. This cluster has
$\beta=0.766$ and $k_BT_g=0.52KeV$ \cite{reiprich}, and the radial
velocity dispersion is about $240km/s$ \cite{adami}. We see that the
radial velocity dispersion (\ref{sigma_5}) is compatible with the
observed profiles \cite{adami, carl}. Since an expression for the
radial velocity dispersion can be obtained by other theoretical
methods which explain cluster discrepancies, such a relation can be
used to differentiate them \cite{harko_virial,bohmer}.

\section{Discussion}
In this paper we have considered the virial mass in the frame work
of DGP brane worlds. The resulting field equations on the brane have
an additional term which is due to the geometry of the extra
dimension and can be associated with a geometrical mass. The virial
theorem was obtained by the use of Boltzmann equation, assuming that
galaxies in the cluster are point-like, non-interacting particles.
We showed that the resulting virial theorem has an additional
potential term due to the extra dimension. The virial theorem has
also been exploited in other brane-world models
\cite{harko_virial,heydari-sepangi} to explain the virial mass
discrepancy. The advantage of the DGP model however is in the
explanation of the self accelerating phase of the universe without
resorting to dark energy in a consistent manner. We obtained the
virial mass of clusters from the virial theorem and showed that it
is proportional to the geometrical mass of the model. The behavior
of the virial mass was investigated in section 4 and showed to be a
linear function of the distance. To estimate the geometrical mass
$\mathcal{N}$ we needed another relation in addition to the Einstein
field equations to close the system of equations. Such a relation
was obtained by using the Jean's equation for clusters, assuming
that clusters have spherical symmetry and are in thermodynamic
equilibrium.

The solution presented in this work offers a possible explanation to
the question of the virial mass discrepancy. The radial velocity
dispersion profile of clusters was also obtained, having two
arbitrary constants, one with a coefficient decreasing with $r$ and
the other increasing with $r$. This allowed us to use this
expression for any cluster with $\beta\geq \frac{2}{3}$. We used the
observed radial velocity dispersion for the cluster NGC5813 as an
example to show that our model can account for the velocity
dispersion of clusters. In addition to explaining the observational
data, the velocity dispersion profile can be used to study the
various aspects of models predicting the virial mass.


\end{document}